\documentstyle[12pt,epsfig]{article}
\newcommand{\be}{\begin{equation}}
\newcommand{\ee}{\end{equation}}
\newcommand{\beqn}{\begin{eqnarray}}
\newcommand{\eeqn}{\end{eqnarray}}

\newcounter{savefig}
\newcommand{\alphfig}{\setcounter{savefig}{\value{figure}}%
\setcounter{figure}{0}%
\renewcommand{\thefigure}{\mbox{\arabic{savefig}\alph{figure}}}}

\begin{document}
\begin{center}
\begin{Large}
{\bf How Does the BFKL Pomeron Couple to Quarks?}\\
\end{Large}
\vspace{0.5cm}
J. Bartels$^a$, J.R.Forshaw$^b$, H.Lotter$^a$,
L.N.Lipatov$^{c\; d}$ \footnote{Alexander von Humboldt Preistr\"ager},
                       M.G.Ryskin$^d$, M.W\"usthoff$^a$ \\
\end{center}
\vspace{0.5cm}
$^a${\it II. Institut f\" ur Theoretische Physik,
Universit\" at Hamburg.}
\\
$^b${\it Rutherford Appleton Laboratory, Chilton, Didcot OX11 0QX,
England.}
\\
$^c${\it DESY, Inst.f\"ur Hochenergiephysik Zeuthen,
Platanenallee 6, 01615 Zeuthen.}
\\
$^d${\it St.Petersburg Nuclear Physics Institute, 188350, Gatchina,
Russia}
\vspace{2.0cm}

{\bf Abstract:} We investigate the coupling of the BFKL Pomeron to
quarks and to colorless states. Special emphasis is given to the
effective quark-quark scattering amplitude of Mueller and Tang.
\vspace{2cm}

1. The BFKL Pomeron \cite{BFKL} is now intensively used for the QCD
description of small-x physics investigated at HERA and other
laboratories. Whereas in many applications (e.g. the deep inelastic
structure function at small x) the solution to the
BFKL equation is needed only for vanishing momentum transfer $t$, in
some particular final states also the nonforward direction $t \neq 0$
enters. For this more general case the solution has been found in
{}~\cite{Lev}, and in its derivation the conformal symmetry of the BFKL
equation plays the key role. This conformal BFKL solution
contains, in momentum space, $\delta$-function like terms which
have no obvious connection with perturbation theory: a physical
interpretation therefore seems to be obscure.

More recently Mueller and Tang~\cite{MT} suggested to define a
quark-quark scattering amplitude at large momentum transfer by
subtracting, from the Lipatov
solution, just these $\delta$-function pieces: it was argued that only
after such a subtraction the nonforward BFKL Pomeron looks like a
result of QCD perturbation theory and thus allows an interpretation
in this language.

So the situation of the general (i.e. nonforward) BFKL solution
looks somewhat unsatisfactory: from the point of view of the conformal
symmetry, it seems doubtful whether the Mueller-Tang prescription
(which is not conformally invariant) represents a solution to the
BFKL equation. From a more intuitive point of view, on the other hand,
one would like to
have a physical interpretation of the conformal solution~\cite{Lev},
e.g. to understand how the impulse approximation emerges. An obvious
way to ``bridge the gap`` between these two approaches is a study of
physical processes, i.e. the scattering of colorless physical states.
In what cases does the quark-quark scattering amplitude with the
modified (\`{a} la Mueller and Tang) Pomeron lead to the same answer as
the conformal solution of~\cite{Lev}? And if so, how are the
$\delta$-function pieces of the conformal solution connected with the
impulse approximation?

In this note we shall try to answer some of these questions by studying
two examples of scattering of colorless states at large momentum
transfer. We begin by adopting the conformal point of view and
briefly reviewing the conformal solution~\cite{Lev}
and the Mueller-Tang~\cite{MT} prescription. We then discuss, using
a more intuitive language, two examples of the scattering of colorless
states at large momentum transfer, and we develop some understanding of
when the quark-quark scattering amplitude of Mueller and Tang will give
the correct answer. Returning to more formal arguments, we then
demonstrate in more detail how the impulse approximation arises from
the conformal solution.
In the final part we say a few words on how the absence of infrared
divergencies in the Mueller-Tang amplitude can be understood.

2. The solution of the homogeneous Bethe-Salpeter equation for the
pomeron wave functon for arbitrary momentum transfer $t=-Q^2$ has the
form ~\cite{Lev}:
\beqn
\Psi_{n,\nu}(k_t,Q-k_t)= \hspace{5cm} \nonumber \\
\int d^2\rho_1d^2\rho_2\exp{i(k_t\rho_{10}+(Q-k_t)\rho_
{20})}\left(\frac{\rho_{12}}{\rho_{10}\rho_{20}}\right)^{\frac 12 +i\nu+\frac
n2}\left(\frac{\rho^*_{12}}{\rho^*_{10}\rho^*_{20}}\right)^
{\frac 12 +i\nu-\frac n2}
\eeqn
where $\rho_{ij}=\rho_i-\rho_j$, $\rho_1,\;\rho_2,\;\rho_0$ are the
complex coordinates of the two gluons and of the pomeron in the
two-dimensional transverse subspace, resp.  $\gamma=\frac{1}{2}+i\nu$
is the anomalous dimension of the composite operator
representing the pomeron, and $n$ denotes its conformal spin. Both
$\gamma$ and $n$ are the quantum numbers of the corresponding
irreducible representation of the conformal group which is the
invariance group of the BFKL equation\cite{Lev}. For unitary
representations $\nu$ and $n$ take real and integer values, resp.

Some time ago Mueller and Tang~\cite{MT} suggested that this expression
(1) might be used for calculating quark-quark scattering amplitude at
non-zero momentum transfer $Q=\sqrt{-t}$. Their arguments were the
following. In order to obtain the coupling of the BFKL
pomeron to a quark one should simply multiply $\Psi_{n,\nu}$ in (1)
with a constant (which represents the quark impact factor)
and integrate the result over $k_t$. As it is seen from eq.(1),
after the integration we obtain a $\delta^{(2)}(\rho_{12})$-function,
and therefore, formally, the pomeron does not couple to a quark.
But Mueller and Tang argued, that in\cite{Lev} the
conformally-invariant solution (1) of the BFKL equation had been
obtained under the assumption that it will be used only for the
scattering of colourless particles.
Therefore in the case of the quark-quark scattering generally it is not
valid. Furthermore Mueller and Tang
noticed that the expression (1) contains terms proportional to
$\delta^{(2)}(k_t)$ and $\delta^{(2)}(Q-k_t)$, as a result of the bad
behaviour of the integrand at $\rho_1\to\infty$ and $\rho_2\to\infty$,
resp. These terms give vanishing contributions when the scattering of
the colourless objects is considered. Therefore there seems to be
some freedom of adding or subtracting such $\delta$-function
terms, and Mueller and Tang suggested to use instead of eq.(1) the
following expression (for simplicity we put $n=0$):
$$\Psi^{MT}_{0,\nu}(k_t,Q-k_t)=
\int d^2\rho_1d^2\rho_2\exp{i(k_t\rho_{10}+(Q-k_t)\rho_{20})}\cdot$$
\begin{equation}
\cdot\left[ \left|\frac{\rho_{12}}{\rho_{10}\rho_{20}}\right|^{1 +2i\nu}\right.
-\left|\frac 1{\rho_{20}}\right|^{1+2i\nu}
\left.-\left|\frac 1{\rho_{10}}\right|^{1+2i\nu}\right].
\end{equation}
which completely removes the singular behavior in (1). Mueller and
Tang argued that such a prescription is closest to perturbation theory
since there are no $\delta$-function pieces coming from
the usual rules. As a result of these changes, the coupling of the
pomeron to a single quark line is no longer zero:
$$\int \frac{d^2k_t}{(2\pi)^2}\Psi^{MT}_\nu(k_t,Q-k_t)=$$
\begin{equation}
=\;-\int d^2\rho_1d^2\rho_2
\left(\left|\frac 1{\rho_{10}}\right|^{1+2i\nu}\right.
+\left.\left|\frac 1{\rho_{20}}\right|^{1+2i\nu}\right)e^{iQ\rho_{20}}
\delta^{(2)}(\rho_{12})
\end{equation}
$$=\;-\pi\;\left(\frac{Q}{2} \right)^{-1+2i\nu}\;
\frac{\Gamma(\frac{1}{2}-i\nu)}{\Gamma(\frac{1}{2}+i\nu)}$$

Bearing in mind that it was the conformal symmetry of the BFKL kernel
which led to the solution (1), the subtraction procedure (2) looks
somewhat strange. Indeed,
so far nobody has proven that both $\Psi_\nu$ and $\Psi^{MT}_\nu$ are
solutions of the BFKL equation with the same eigenvalue $\omega$, and
therefore the conformal invariant expression (1) is preferable. Of
course,
the BFKL equation may have non-conformal invariant solutions. We know
only that the set of the conformally invariant solutions is complete in
the space of the generalized functions, which are integrated with
impact factors with some good properties. Only impact factors of
colourless particles have these properties. Further, the presence of
$\delta$ functions in the solution of homogeneous BFKL equation
generally
does not contradict perturbation theory. One should
find the solution of the inhomogeneous equation by expanding it in the
series over 'singular' functions (1), as it was done in\cite{Lev}, and
verify that up to the terms which give a zero contribution for the
colourless particle scattering the result is in agreement with the
perturbation theory (cf. the appendix of ref.\cite{Lev2}).

The appearence of the singular ($\delta$-function) term in the
solution (1) is related to the fact that in the leading logarithmmic
($\ln\frac 1x$) approximation the anomalous dimension $\Delta$
of the gluonic
field $\phi (x_i)$ in the Polyakov ansatz \cite{Pol} for the three point
function
\beqn
E_{n, \nu} (\rho_{10}, \rho_{20}) & = &
<\phi (\rho_1)\phi (\rho_2)O_{n,\nu}(\rho_0)>    \nonumber \\
                                  & = &
|\rho_{12}|^{-2\Delta} \left(\frac{\rho_{12}}{\rho_{10}\rho_{20}}\right)
^{\frac 12 +i\nu+\frac n2} \left(\frac{\rho^*_{12}}{\rho^*_{10}\rho^*_{2
0}}\right)^{\frac 12 +i\nu-\frac n2}
\eeqn
vanishes \footnote{In the leading logarithmic approximation the result
$\Delta=0$ follows from the normalization condition and the hermiticity
property of the BFKL kernel.}.
For $\Delta\neq 0$ eq.(1) contains only
smeared $\delta$-functions, which are not in the contradiction with the
perturbation theory.
In two-dimensional conformal field theories it is natural to
introduce the infinitesimal dimension $\Delta$ for the field $\phi$
as a regulator: its Green's function is proportional to $\ln
|\rho_{12}|$, and it can be obtained from the general conformal
invariant expression $\sim |\rho_{12}|^{-2\Delta}$ by the limiting
procedure $\Delta\to 0$. From this point of view the substitution
$\Psi_{n,\nu}\to \Psi^{MT}_{n,\nu}$ also looks unnatural.

Returning to the quark-quark scattering amplitude, one might adopt the
view that, after all, such an amplitude is not physical, and therefore
one is allowed to select any prescription for this amplitude. The
physical quantity is the scattering amplitude for the colourless
states. One then has to reformulate the problem in terms of
observable scattering amplitudes:
what is the form of the hadron amplitude, in case where the momentum
transfer is significantly bigger than the essential transverse
momenta of the partons inside the colliding particles? One can
expect that in this kinematic region the result will have the form of
the impulse approximation, with the effective quark-quark scattering
amplitude being averaged with the parton wave functions.
Indeed, one of the results obtained in\cite{FR} (and to be
reconsidered further below) confirms the Mueller-Tang
recipe for the effective quark quark scattering: namely starting from
the conformally invariant solution (1) (i.e without any subtraction of
the type of eq.(2)), the authors found, in one of the cases they
investigated, that the result has the form of the impulse approximation
although, formally, only the interaction
with the different quarks ($\rho_1\neq\rho_2$) contributes to the
amplitude. In the large-$t$ limit, the scattering amplitude is
dominated by
the $\delta$ function pieces inside (1), and the leading term
has the same form as one would have obtained under the assumption
that the BFKL pomeron interacts with only one quark line.
On other words, one obtains the same result as given by the Mueller-Tang
prescription, i.e. without taking into account the interaction with
different quark lines. However, one might also expect situations where
the large-$t$ behavior of the scattering amplitude of colorless objects
also feels the nonsingular part of (1): in this case the simultaneous
interaction of the Pomeron with different partons becomes important,
and the Mueller-Tang prescription should not be applied (or has
to be modified).

3. Let us now turn to the examples which are taken from ~\cite{FR}.
It is convenient to use the notations $R=(\rho_1+\rho_2)/2$ and
$\rho=\rho_1-\rho_2$. Let $\Phi (\rho )$ be the wave function of the
initial hadron, where we omit all the arguments (coordinates)
except of $\rho$. As an example, one may think of the scattering of an
onium state where $\rho$ is the separation between the quark and
antiquark in the impact parameter plane.

We have to start from the conformal expression (1), as this is the only
known solution to the BFKL
equation. So we write, as the first example, the hadron-pomeron vertex
$V(Q)$ as the convolution of the eigenfunction
$\Psi_{n,\nu}=\left(\frac{\rho_{12}}{\rho_{10}\rho_{20}}\right)^
{\frac 12 +i\nu+\frac n2}$ with the square of the hadron wave function:
\beqn
V(Q)=\int d^2\rho d^2R\left(\frac{|\rho |}{|R+\frac
{\rho}2||R-\frac{\rho}2| } \right)^{1 +2i\nu}e^
                    {iQR}|\Phi (\rho)|^2
\eeqn
where we have put $n=0$.

As it was stressed before, from the {\it formal} point of view the
expression (5) has to be associated with the graph Fig.1a (i.e. the
interaction with two different quarks): the coupling of $\Psi_{n, \nu}$
to a single parton line gives zero since $\Psi_{n,\nu}=0$ at $\rho=0$.
Nevertheless, in the large $Q$ limit there is a contribution
of eq.(5)) which comes from the points $R\to \pm \rho /2$ where the
variation of the eigenfunction is largest and has the same form as the
interaction with a single parton line (graph Fig.1b).
Indeed, if $\rho >>\Delta R\sim 1/Q$ the integration
near the point $R\to \rho /2$ gives
\beqn
\int  d^2R\left(\frac{|\rho
|}{|R+\frac {\rho}2||R-\frac{\rho}2|} \right)^{1
                                +2i\nu}e^{iQR}
              & \simeq &
\int d^2\rho'\left( \frac 1{\rho'}\right)^{1 +2i\nu}
               e^{iQ\rho/2+iQ\rho'}
\nonumber \\
             & = &
\frac{\pi}{(Q/2)^{1-2i\nu}}e^{iQ\rho
/2}\frac{\Gamma(1/2-i\nu)}{\Gamma(1/2+i\nu)}
\eeqn
The same result would have been obtained if we had used the
Mueller-Tang prescription and taken into account only the diagram
Fig.1b (note that the Mueller-Tang subtraction has a
negative sign. It reflects the opposite sign of the colour coefficients
corresponding to the diagrams Fig.1a and Fig.1b).

The interpretation of this contribution as representing the interaction
of the BFKL Pomeron with a single quark at the point $\rho_1=\rho/2$
is based upon its $Q$ dependence. We take the point of view that the
$Q$-dependence is the best (maybe even the only) way to distinguish
between the contributions of Fig.1a and 1b. For the impulse
approximation
\footnote{By impulse approximation we mean the approximation
in which the interaction a compound system of several partons can
be described as a sum of interactions with each parton seperately.}
(Fig.1b) one expects an expression of the form
\begin{equation}
\int d^2\rho \Phi^*(\rho )e^{iQ\rho /2}\Phi(\rho ) =F(Q),
\end{equation}
where $F(Q)$ is the hadron form factor which vansishes for large $Q$.
In contrast to this, Fig.1a is described by the following function:
\beqn
\int d^2 \rho \Phi^*(\rho) ) e^{i(Q-k)\rho/2 -ik\rho/2} \Phi(\rho),
\eeqn
i.e. the appearance of a $k$-dependent exponential signals that both
the quark and the antiquark share the transverse momentum $Q$.
Our discussion above then implies that the conformal Pomeron (1),
when coupled to (8) and restricted to the region of integration
$R=\pm \rho/2$, produces the exponential $e^{i \pm Q\rho/2}$ and is
interpreted as being associated with graph Fig.1b.

So we can proceed in two equivalent ways:\\
a) either we use the 'conformal' pomeron wave functions $\Psi^{n,\nu}$
(as we have done). In this case the convergence of the $\rho_i$
integrals is provided by the external wave function $\Phi$.
As an additional advantage, this procedure also slightly reduces the
number of graphs (no coupling to the single parton line). But in this
way the physical interpretation in terms of Feynman diagrams becomes
more difficult, and it becomes harder to exploit our physical intuition.
\\
b) Alternatively, we could perform the Mueller-Tang subtractions in
ALL the graphs Figs.1a and b (the sum of all subtraction terms gives
zero due to the colourless of the initial state (hadron)).
Now the integrand falls down with $\rho$, even without invoking any
$\Phi$-function, and we get back the simple physical interpretation of
the Feynman graphs. This method seems to be most useful in the case
when one expects that the dominant contribution comes from the impulse
approximation diagram: then the Mueller-Tang prescription provides a
crucial simplification. The accuracy to which the impulse
approximation may hold is controlled by the parameter
$1/(Q\rho)^2$: if the essential $\rho >>1/Q$ one can neglect the
coupling to the different parton lines; the corrections to
the impulse approximation result are of the order of
$O(\frac1{Q^2\rho^2})$.

So we are lead to ask the question which part of the $\rho$ integration
gives the dominant contribution. In our previous example (5)
the parameter $1/(\rho Q)^2\sim 1$, as the typical $\rho$ in the
integral is of about $1/Q$ due to the exponent $e^{iQ\rho /2}$. Even
more, the contribution which corresponds to the coupling to different
lines (Fig.1a) and comes from the "corrections" $1/(\rho Q)^2$ turns
out to be larger. The explanation is that the
transfered momentum $Q$ can be balanced ($ k\simeq Q/2$),
and one gets a result which does not decrease with $Q$
($k=Q/2 + l$):
\beqn
\int d^2\rho |\Phi (\rho)|^2\exp{(i(Q-k)\rho/2-ik\rho/2)}d^2k
             &\approx &
\int d^2\rho |\Phi (\rho)|^2\exp{(-il\rho)}  \nonumber \\
& =& \int F(2l)d^2l=const
\eeqn
instead of the form factor $F(Q)$ which falls down steeply at large $Q$.
So for the example (5) the impulse approximation does not give the
leading contribution, and the use of the M-T prescription is not very
useful.

However it is possible to change the situation in such a way that
the large $\rho$ region dominates: $\rho \gg 1/Q$.
Instead of the hadron-Pomeron vertex (5)
one has to consider an inclusive process in which
the initial hadron is destroyed (see fig.2). In this case
we get, in the expression for the cross section, two different integrals
in $R$ and $R'$ (one for the amplitude $A$, one for the
complex conjugate $A^*$), but still only one integral over the
parton-parton separation $\rho$. Instead of (5) we now have
\footnote{For symplicity we put here $\nu =0$, which is the
dominant value in the high energy limit.}:
\beqn
I(Q)=\int d^2\rho |\Phi
(\rho)|^2
          \frac{|\rho |}{|R-\rho/2||R+\rho/2|}
          \frac{|\rho |}{|R'-\rho/2||R'+\rho/2|}
                              e^{iQR-iQR'}
\eeqn
The contributions from the "singular" points $R=R'=\pm\rho/2$
(or, in other words, the ``Mueller-Tang terms`` ($-1/\rho_1-1/\rho_2$)
in
the Feynman diagram which describe the pomeron coupling to the single
parton line) now do not have the
``dangerous`` exponents $\exp{(iQ\rho/2)}$. Thanks to their opposite
signs the exponents of $\exp{(iQR)}$ and $\exp{(-iQR')}$  cancel each
other, and the result takes the form:
\beqn
I(Q)\simeq \left(\frac{2\pi}Q\right)^2 2\int d^2\rho|\Phi (\rho )|^2
\eeqn
(the factor 2 counts the two contributions from the points
$R=R'=-\rho/2$ and $R=R'=\rho/2$). In the graph of Fig.2 the lower part
represents, at large $Q$,
the structure function (say, $xG(x,Q)$) which in the LLA contains the
logarithm $\int^{Q^2}_{\mu^2}d^2q/q^2$. In the $\rho$ representation it
means that $|\Phi (\rho )|^2\propto 1/\rho^2$ and the integral (11) over
$\rho$ takes the form  $\int^{1/\mu}_{1/Q} d^2\rho/\rho^2$ where the
typical $\rho\sim \sqrt{1/Q\mu}>>1/Q$. So the essential values of
$\rho$ in (11) are large, and the correction $\sim 1/(\rho Q)^2$ is
small indeed.

Thus we presented an example where the Mueller-Tang prescription
works. The criterion is the following. If the distance between the
"active" (i.e. interacting with the pomeron) quark and the spectators is
large in comparison with the inverse momentum transfer, the
M-T prescription can be used. Such a condition can be fulfilled in
inclusive processes, but not
in the exclusive scattering (elastic, photoproduction or
electroproduction of a vector meson). In the latter case the
separation between the quarks is determined by the
momentum transfer $Q$; the M-T prescription has to be generalized as
stated above, and it does not simplify the
calculations. The coupling to different parton lines becomes
important, and method (a) involving the 'conformal' pomeron function
needs fewer diagrams. In any case, one has to check
whether the ``distance`` criterion is satisfied or not.

4. Let us take a closer look at the conformal Pomeron and see how
it manages to reproduce the impulse approximation. We consider
the same examples as before, i.e. the scattering of colorless
states of two quarks in the large-$t$ limit. To be definite,
let us start with the coupling of the pomeron to the $J/\Psi$
production vertex with a virtual photon as initial particle (figs.1a and 1b).
The momentum transfer $t$ is assumed to be fixed and large compared to
the virtuality of the photon $Q_{\gamma}^2$ and the mass of the $J/\Psi$,
 $M^2$.
Due to the large mass of the heavy quarks the nonrelativistic
approximation of the meson wave function can be applied which leads to
a simple formfactor at the upper photon-meson vertex \cite{FR}.
The virtual photon dissociates is a quark antiquark system
which can be represented as a wave function depending on the distance
$\rho=\rho_1-\rho_2$ of the two quarks.
Instead of eq.(1) we use the mixed representation
\be
\Psi_{0,\nu}(\rho,Q)=\int d^2\rho_2 \frac{\rho^{1+2i\nu}}{\left[(\rho+\rho_2)^2
\rho_2^2\right]^{\frac{1}{2}+i\nu}}\;e^{iQ\rho_2}
\ee
The wave function of the quark-antiquark system was found to be
$K_0(Q_{||}\rho) e^{i Q/2 \cdot \rho}$
($Q_{||}^2 = (Q_\gamma^2+M^2)/4$, see \cite{FR}),
and the analytic expression of the production vertex is:
\be
\int d^2 \rho \, K_0(Q_{||}\rho)e^{i Q/2 \cdot \rho}\;
\Psi_{0,\nu}(\rho,Q) \;\;.
\ee
Only fig. 1a contributes whereas 1b is zero as was already discussed before.
Mainly due to the phase factor $e^{i Q/2 \cdot \rho}$ in eq (6),
which is a consequence of the bound
final state and which leads to a formfactor, a simple
factorization of the wave function is impossible and the Mueller
Tang prescription fails. We insert the Mellin transformed of $K_0$
with $\lambda$ as a real variable and derive from eq.(6) the following
expression:
\beqn&&
4\pi^4 \;\frac{(Q/2)^{-3+2i\nu_1}}{\Gamma^2(\frac{1}{2}+i\nu)}
\;\int_{-\infty}^{\infty} \frac{d\lambda}{\pi} \,\left(\frac{Q^2}{4Q_{||}^2}
\right)^{3/2+i\lambda} \;\Gamma^2(\frac{3}{2}+i\lambda)\nonumber
\\ && \frac{\Gamma(\frac{1}{2}-i\nu)\Gamma(\frac{1}{2}+i\nu)
\Gamma(-i\lambda-i\nu)\Gamma(-i\lambda+i\nu)}
{\Gamma \left(1+\frac{i\lambda+i\nu}{2}\right)
\Gamma \left(1+\frac{i\lambda-i\nu}{2}\right)
\Gamma \left(1-\frac{i\lambda+i\nu}{2}\right)
\Gamma \left(1-\frac{i\lambda-i\nu}{2}\right)}\;\;.
\eeqn

Having a closer look at eq.(7) we find a double pole in the complex
$\lambda$-plane at $\lambda=3/2i$. This double pole leads to a
log($Q^2/Q_{||}^2$) in addition to the basical power
behaviour $Q^{-3}=(-t)^{-3/2}$. A more detailed analysis shows that the
momentum distribution of the two gluons is symmetric, i.e each gluon carries
roughly a half of the total momentum $Q$.
Moving the contour of integration
beyond this pole we collect all the nonleading contributions which contain
extra powers in $Q_{||}^2/Q^2$.

The second case concerns the coupling of two pomerons (conformal
dimensions $\nu_1$ and $\nu_2$) to one quark-antiquark system which
originates from the onium state (fig.2c). The smallest scale $Q_0^2$ is
given by the size of the onium and is supposed to be much smaller than
$-t$. This configuration corresponds to the onium dissociation (figs.2a
 and 2b) into an open quark-antiquark system. The analytical expression
to be calculated is (see also ref.\cite{BLW}):
\be
\int d^2 \rho\, \left|\Phi_{onium}(\rho)\right|^2  \;
\Psi_{0,\nu_1}(\rho,Q)  \Psi_{0,\nu_2}(\rho,-Q)\;\;.
\ee
For similar reasons as in eq.(6) only fig.2c contributes.
We remark that in eq.(8) an additional phasefactor is absent in
contrast to eq.(6). Inserting the Mellin transform of the
onium wave function analogous to the treatment of the
$K_0$-Bessel function in eq.(6)
\be
\int_0^{\infty} d\rho \rho\; \left|\Phi_{onium}(\rho)\right|^2
\;\rho^{1+2i\lambda} \;\;,
\ee
we are able to factorize off the wave function. The remaining
contribution from eq.(9) is:
\be
\int_0^{\infty} d\rho \rho^{-2-2i\lambda}\;\int_0^{2\pi}d\phi\;
\Psi_{0,\nu_1}(\rho, Q)  \Psi_{0,\nu_2}(\rho,-Q)\;\;.
\ee
The onium wave function may contain UV-singularities, especially if one
takes into
account radiative corrections (a multiple pole occurs at $\lambda=i/2$):
the factorization has to be performed in such a way that all (collinear-)
singularities including  those of the vertex (eq.(8)) are absorbed
into the onium structure function. This
procedure is very similar to the usual 'Mass factorization' where
$\lambda$ is the dimensional regulator. To be more specific let us
assume that the onium is a virtual photon with the scale $Q_0^2$. The
photon wave function is singular at $\rho=0$ and produces one pole.
Comparing to the usual deep inelastic scattering we know that the
photon  structure function at lowest order perturbation theory has at
least one logarithm due to the point like structure of the photon. A
logarithm in $Q^2/Q_0^2$ requires at least a double pole in $\lambda$
at the point $\lambda=i/2$, i.e if factorization holds, we except one
more pole in expression (8).

In order to evaluate expression (8) we take the Mellin transformation
of both pomeron wave functions introducing the variables ($\lambda_1$
and $\lambda_2$) and find:
\beqn &&
\frac{\pi^2 \sqrt{\pi}}{4\;\Gamma(\frac{1}{2}+i\lambda)} \;
\frac{2^{-2i\lambda}\left(\frac{Q}{2}\right)^{-1+2i\nu_1+2i\nu_2+2i\lambda}}
{\Gamma^2(\frac{1}{2}+i\nu_1)\Gamma^2(\frac{1}{2}+i\nu_2)}\;
\;\int_{-\infty}^{\infty}\frac{d\lambda_1}{\pi}
\int_{-\infty}^{\infty}\frac{d\lambda_2}{\pi}
\Gamma(-i\lambda_1+i\nu_1) \nonumber \\ &&
\Gamma(-i\lambda_1-i\nu_1)
\Gamma(-i\lambda_2+i\nu_2) \Gamma(-i\lambda_2-i\nu_2)\;
\frac{\Gamma(\frac{1}{2}-i\lambda+i\lambda_1+i\lambda_2)}
{\Gamma(\frac{1}{2}+i\lambda-i\lambda_1-i\lambda_2)}  \\ \nonumber &&
\Gamma(i\lambda-i\lambda_1-i\lambda_2)\;
\frac{\Gamma(\frac{1}{2}+i\lambda_1)\Gamma(\frac{1}{2}+i\lambda_2)
\Gamma(2i\lambda-i\lambda_1-i\lambda_2)}
{\Gamma(\frac{1}{2}+i\lambda-i\lambda_1)
\Gamma(\frac{1}{2}+i\lambda-i\lambda_2)}\;\;.
\eeqn
A careful investigation of expression (11) shows that a double
pinching in $\lambda_1$ and $\lambda_2$ occurs as soon as $\lambda$
approaches $i/2$, i.e the integrals over $\lambda_1$ and $\lambda_2$
induces a double pole in $\lambda$ at $\lambda=i/2$. The
$\Gamma(1/2+i\lambda)$ in the denominator in front of
expression (11) reduces the double pole to a single pole.
The residue is simply the factor
\be
\pi^2\;\left(\frac{Q}{2}\right)^{-2+2i\nu_1+2i\nu_2}\;
\frac{\Gamma(\frac{1}{2}-i\nu_1)}{\Gamma(\frac{1}{2}+i\nu_1)}
\frac{\Gamma(\frac{1}{2}-i\nu_2)}{\Gamma(\frac{1}{2}+i\nu_2)}\;\;.
\ee
which is identical to the expression resulting from eq. (3).
In addition, a logarithm in $Q^2/Q^2_0$ occurs reflecting the pointlike
structure of the photon. It has to be absorbed in the photon wave
function. In the case of a nonsingular onium wave function we get an
overall factor $\int d^2 {\bf\rho}\left|\Phi_{onium}\right|^2$.

The result (19) agrees with what one would have obtained by using only
the $\delta$-function pieces of the pomeron wave function. We therefore
conclude that the large-$t$ limit is governed by these singular pieces;
correspondingly, the distribution
of the momenta of the two gluons in the conformal pomeron is very
asymmetric: one gluon is very soft whereas the other one carries the
total momentum $Q$. Why do we, nevertheless, get the same result as in
in the impulse approximation? The answer is very simple: the soft gluon
line of the conformal Pomeron does not distinguish to which of the
quark lines it couples. The coefficient of the $\delta$ function, on
the other hand, is identical to the result from the impulse
approximation. The result therefore looks identical to the impulse
approximation and confirms the Mueller-Tang prescription.

5. The last question we would like to discuss is the infrared logarithm
which is present in the two gluon exchange quark-quark amplitude (fig.3)
but disappears in the asymptotic Mueller-Tang formula\cite{MT}.
Why does the small $k_t$ (see fig.3) region do not contribute to the high
energy amplitude? The answer to this question was given in
ref.\cite{FR}: beginning at small energies, one can investigate,
step by step, the disappearence of the infrared logarithm. As long as
$z=\frac{N_c\alpha_s}{2\pi}\ln\frac s{Q^2}<1$ it is possible to
sum up the double-log terms of the type $(\alpha_s\ln\frac
s{Q^2}\ln\frac{Q^2}{k^2_t})^n$. These terms comes from the
reggeizized part of the BFKL kernal\cite{BFKL}, i.e. from the Feynman
graphs fig.3b,c. They are easily exponentiated ($z\leq 1$ and
$k<<Q$):
\begin{equation}
f(k;Q,z)\propto\frac{(k^2_t/Q^2)^z}{k^2_t(Q-k)^2_t}
\end{equation}
As a result of this exponentiation, the contribution of the region
$k_t<<Q$ dies out with energy, at least up to $z\leq1$. At larger
energies this simple formula can no longer be used since subdominant
terms become important which have been neglected in (7). For $z>1$
nothing dramatical happens near $k_t=0$, and when iterating the
integral equation for the BFKL Pomeron
the behaviour of the n-th iteration $f_n(k;Q,z)$ at small $k<<Q$
is determined mainly by the value of the previous iteration  step
$f_{n-1}(k;Q,z)$ in the region $k\sim Q/2$.
To see this absence of infrared logarithms most clearly we return to
the $\rho$-representation and subtract from Lipatov's conformal solution
(1) the $\delta(k_t)$ term, i.e. we make use of the Mueller Tang
formula (2). As a result of the subtractions, infrared divergencies are
absent. In the high energy limit typical values of $\nu$ are small
($|\nu |<1/2$). In the limit $Q>>k$ the essential $\rho_2$-values are
of the
order $\sim 1/Q$ and, after the subtraction, the essential region
of $\rho_1$ is of the order of
$\rho_2$; i.e. $\rho_1\sim 1/Q$ too. This means that
in the region $k_t<<Q$ the amplitude (2) becomes practically
independent of $k_t$. In particular it does not contain the singularity
$1/k_t^2$ which in the two gluon exchange diagram 3a
leads to the infrared logarithm $\int^{Q^2}_{k_0^2}d^2k/
k^2=\ln Q^2/k_0^2$ ($k_0$ is the infrared cutoff).

The situation changes crucially in the case of $\nu\to i/2$,  which
corresponds to the "low" energy (DGLAP) limit. From the formal point of
view the expression in the square brackets still goes to zero as
$\rho_1\to\infty$, but only slowly. Therefore we have no
$\delta$-functions ($\delta(k_t)$) in eq.(2). But
when the power $1/2+i\nu$ becomes very small the square bracket
is close to -1 over a very large region of $\rho_1$, and when
integrating over this region one gets
$f(k;Q)\propto\frac 1{k^2(Q-k)^2}$, as it should be for the two gluon
exchange amplitude. This example shows how in eq.(2) the infrared
logarithm (i.e. the behaviour $f(k,Q)\propto 1/k^2$) is restored in the
DGLAP limit $(\nu\to i/2$).

At first sight this calculation seems to give the wrong minus sign.
However, this sign corresponds to the diagram fig.1a, where the two
t-channel gluons couple to different quarks. On the other hand we
know that the infrared logarithm comes from the Feynman graph Fig.1b
where both gluons interact with the same coloured parton. Since both
quarks form a colourless state the colour coefficient in Fig.1b
has the opposite sign compared to Fig.1a, and the minus sign in the
square brackets of eq.(8) is just a result of this color coefficient.

We hope that this discussion gives a better understanding of
the conformal BFKL pomeron eigenfunctions, helps to clarify the origin
of the Mueller-Tang prescription and, in particular, provides
a criterium for its use in describing the high energy scattering of
colorless states.

{\bf Acknowledgements:} L.Lipatov and M.Ryskin wish to thank DESY and
the II.Institut f\"ur Theoretische Physik, Universit\"at Hamburg,
for their hospitality. \\ \\
This work has been supported in part by
the Alexander von Humboldt Foundation (L.N.Lipatov), the Volkswagen
Stiftung (M.G.Ryskin), and the Deutsche Akademische Austauschdienst
DAAD (J.R.Forshaw).

\section*{Figure captions}
\begin{description}
\item Fig. 1a  : Graphical representation of the coupling of the
                 conformal covariant three-point function
                 $\Psi_{0,\nu}$ to the Photon-Meson vertex.
                 Here the gluons represented by the wavy lines
                 couple to different quarks of the
                 meson wavefunction.
\item Fig. 1b  : The same as in fig.1a but with the two gluons
                 coupling to the same quark of the meson
                 wavefunction.
\item Fig. 2a  : Coupling of the conformal covariant three-point
                 function to the quark-antiquark system originating
                 from an onium state. Gluons are coupled to
                 different quarks of the onium wavefunction.
\item Fig. 2b  : Same as in fig.2a but with the two gluons coupling
                 to the same quark of the onium wavefunction.
\item Fig. 2c  : Graphical representation of the cross section of the
                 dissociation of an onium state into an open
                 quark-antiquark state (cf. eqs. (10),(15)).
\item Fig. 3a  : The two-gluon exchange quark-quark-amplitude.
\item Fig. 3b  : Higher order contribution to the quark-quark
                 amplitude which is part of the reggeized part of
                 the BFKL-kernel.
\item Fig. 3c  : Another contribution to the reggeized part of
                 the BFKL-kernel.
\end{description}
\newpage
\setlength{\unitlength}{1cm}
\setcounter{figure}{1}
\alphfig
\begin{figure}[t]
\begin{center}
\begin{picture}(8,8)(0,0)
%\put(0,0){\framebox(8,8)}
\put(1,0){\epsfig{file=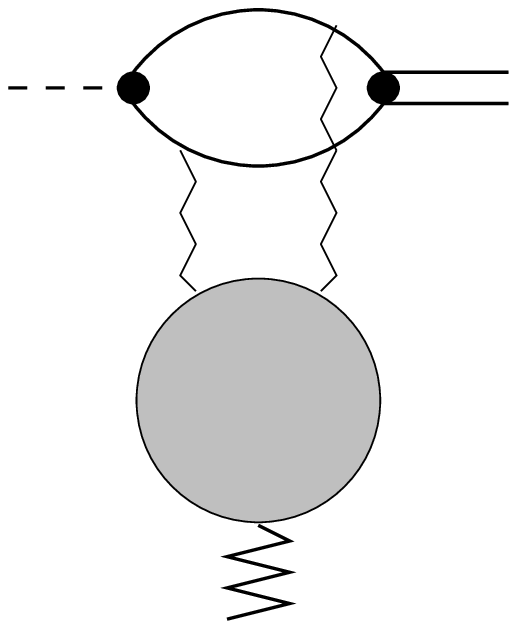,height=8cm,width=6cm}}
\put(3.8,2.5){${\Psi_{0,\nu}}$}
\put(3.0,0.4){$Q$}
\put(5.3,5.0){$Q-k_t$}
\put(2.4,5.0){$k_t$}
\put(3.05,6.2){\circle*{0.1}}
\put(4.9,7.75){\circle*{0.1}}
\end{picture}
\caption{}
\label{fig1a}
\end{center}
\end{figure}
\begin{figure}[b]
\begin{center}
\begin{picture}(8,8)(0,0)
%
%\put(0,0){\framebox(8,8)}
\put(1,0){\epsfig{file=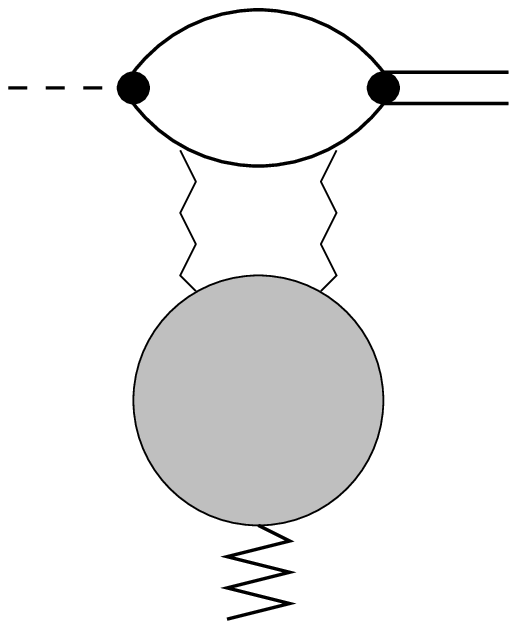,height=8cm,width=6cm}}
\put(3.8,2.5){${ \Psi_{0, \nu}}$}
\put(3.05,6.2){\circle*{0.1}}
\put(4.9,6.2){\circle*{0.1}}
\put(3.0,0.4){$Q$}
\put(5.3,5.0){$Q-k_t$}
\put(2.4,5.0){$k_t$}
\end{picture}
\caption{}
\label{fig1b}
\end{center}
\end{figure}
\setcounter{figure}{2}
\alphfig
\begin{figure}[t]
\begin{center}
\begin{picture}(8,8)(0,0)
%\put(0,0){\framebox(8,8)}
\put(0,0){\epsfig{file=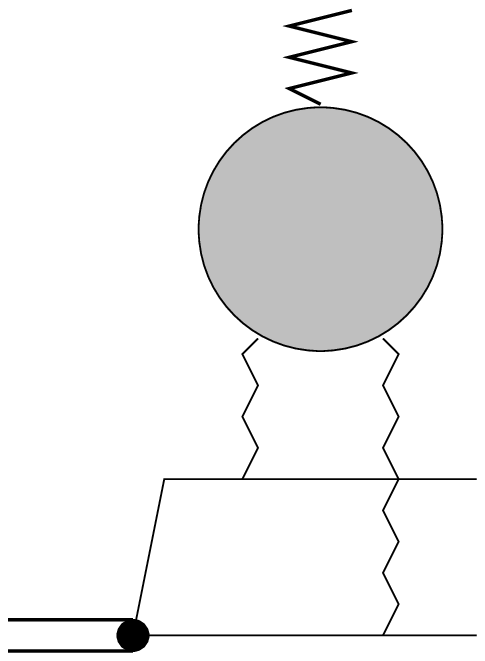,height=8cm,width=6cm}}
\put(2.95,2.15){\circle*{0.1}}
\put(4.75,0.2){\circle*{0.1}}
\put(3.0,7.4){$Q$}
\put(3.8,5){${ \Psi_{0,\nu}}$}
\put(5.3,3.0){$Q-k_t$}
\put(2.4,3.0){$k_t$}
\end{picture}
\caption{}
\label{fig2a}
\end{center}
\end{figure}
\begin{figure}[b]
\begin{center}
\begin{picture}(8,8)(0,0)
%\put(0,0){\framebox(8,8)}
\put(0,0){\epsfig{file=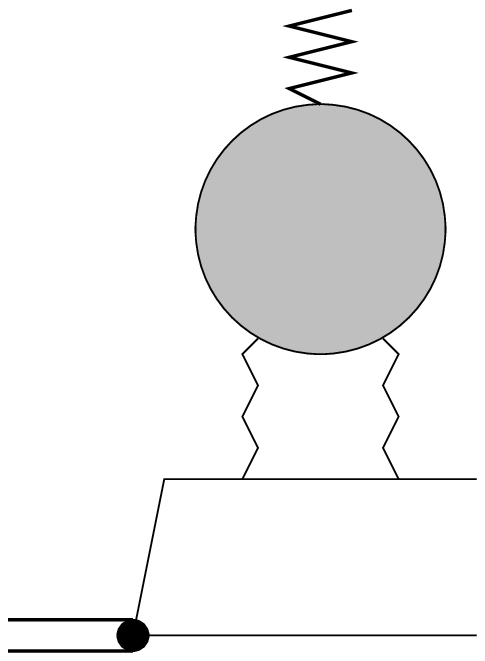,height=8cm,width=6cm}}
\put(2.95,2.15){\circle*{0.1}}
\put(4.9,2.15){\circle*{0.1}}
\put(3.0,7.4){$Q$}
\put(3.8,5){${ \Psi_{0,\nu}}$}
\put(5.3,3.0){$Q-k_t$}
\put(2.4,3.0){$k_t$}
\end{picture}
\caption{}
\label{fig2b}
\end{center}
\end{figure}
\begin{figure}[t]
\begin{center}
\begin{picture}(8,8)(0,0)
%\put(0,0){\framebox(8,8)}
\put(-2,0){\epsfig{file=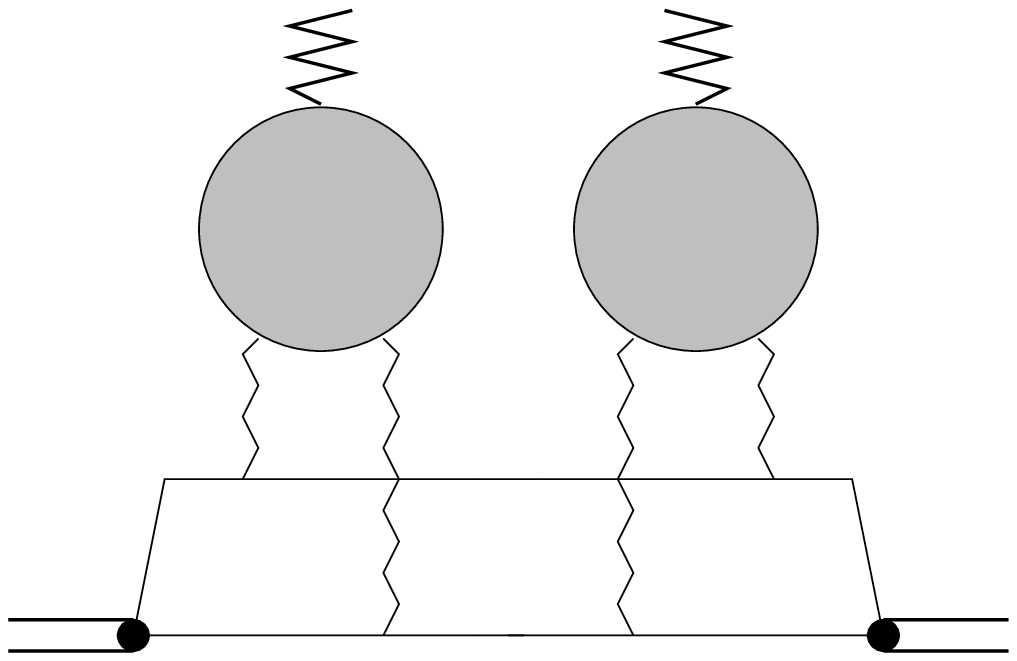,height=8cm,width=12cm}}
\put(0.75,2.15){\circle*{0.1}}
\put(2.45,0.2){\circle*{0.1}}
\put(7.1,2.15){\circle*{0.1}}
\put(5.45,0.2){\circle*{0.1}}
\put(1.4,5){${ \Psi_{0,\nu_1}}$}
\put(6.0,5){${ \Psi_{0,\nu_2}}$}
\put(0.8,7.4){$Q$}
\put(5.0,7.4){$-Q$}
\put(3.0,2.8){$Q-k_t$}
\put(0.4,2.8){$k_t$}
\put(7.5,2.8){$-Q-l_t$}
\put(4.9,2.8){$l_t$}
\end{picture}
\caption{}
\label{fig2c}
\end{center}
\end{figure}
\setcounter{figure}{3}
\alphfig
%\begin{figure}[b]
%\begin{center}
%\begin{picture}(8,8)(0,0)
%\put(0,0){\framebox(8,8)}
%\put(0,0){\epsfig{file=rys3a.ps,height=4cm,width=4cm}}
%\end{picture}
%\caption{}
%\label{fig3a}
%\end{center}
%\end{figure}
%
%
%\begin{figure}[t]
%\begin{center}
%\begin{picture}(8,8)(0,0)
%\put(0,0){\framebox(8,8)}
%\epsfig{file=rys3b.ps,height=4cm,width=4cm}
%\end{picture}
%\caption{}
%\label{fig3b}
%\end{center}
%\end{figure}
%
%
%\begin{figure}[t]
%\begin{center}
%\begin{picture}(8,8)(0,0)
%\put(0,0){\framebox(8,8)}
%\put(0,0){\epsfig{file=rys3c.ps,height=4cm,width=4cm}}
%\end{picture}
%\caption{}
%\label{fig3c}
%\end{center}
%\end{figure}
%
\begin{figure}[b]
\begin{minipage}[t]{5cm}
\begin{picture}(5,5)(0,0)
%\put(0,0){\framebox(5,5)}
\put(0.5,0){\epsfig{file=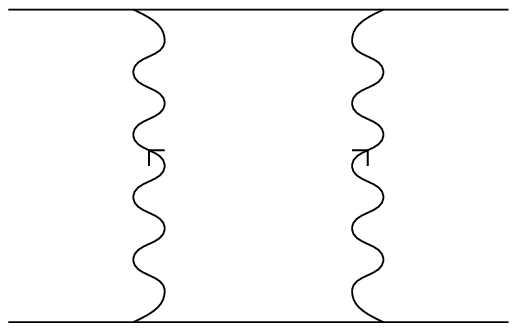,height=4cm,width=4cm}}
\put(1.5,0.05){\circle*{0.1}}
\put(3.5,0.05){\circle*{0.1}}
\put(1.5,4.0){\circle*{0.1}}
\put(3.5,4.0){\circle*{0.1}}
\put(1.0,1.8){$k_t$}
\put(3.7,1.8){$Q-k_t$}
\end{picture}
\caption{}
\end{minipage}
\hfill
\begin{minipage}[t]{5cm}
\begin{picture}(5,5)(0,0)
%\put(0,0){\framebox(5,5)}
\put(0.5,0){\epsfig{file=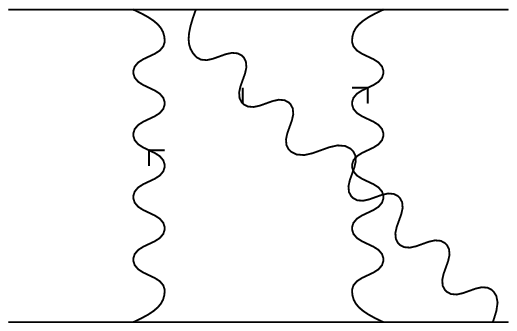,height=4cm,width=4cm}}
\put(1.5,0.05){\circle*{0.1}}
\put(3.5,0.05){\circle*{0.1}}
\put(1.5,4.0){\circle*{0.1}}
\put(3.5,4.0){\circle*{0.1}}
\put(4.35,0.05){\circle*{0.1}}
\put(1.95,4.0){\circle*{0.1}}
\end{picture}
\caption{}
\end{minipage}
\hfill
\begin{minipage}[t]{5cm}
\begin{picture}(5,5)(0,0)
%\put(0,0){\framebox(5,5)}
\put(0.5,0){\epsfig{file=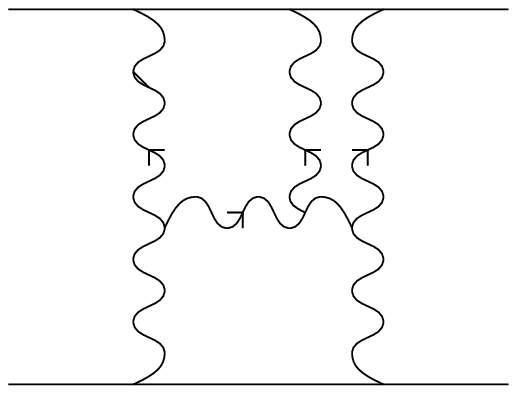,height=4cm,width=4cm}}
\put(1.5,0.05){\circle*{0.1}}
\put(3.5,0.05){\circle*{0.1}}
\put(1.5,4.0){\circle*{0.1}}
\put(3.5,4.0){\circle*{0.1}}
\put(1.7,1.7){\circle*{0.1}}
\put(3.2,1.7){\circle*{0.1}}
\put(2.85,1.85){\circle*{0.1}}
\put(2.7,4.0){\circle*{0.1}}
\end{picture}
\caption{}
\end{minipage}
\end{figure}

\end{document}